
\input harvmac

\noblackbox

\Title{\vtop{\baselineskip=12pt\hbox{IFP-432-UNC}\hbox{hep-th/9205094}}}
{\vbox{\centerline{Running Gauge Couplings and Thresholds}
\vskip2pt\centerline{in the Type~II Superstring}}}
\baselineskip=16pt
\smallskip
\centerline{L.~Dolan and James T.~Liu}
\bigskip\centerline{\it Institute of Field Physics}
\centerline{\it Department of Physics and Astronomy}
\centerline{\it University of North Carolina}
\centerline{\it Chapel Hill, NC 27599-3255, USA}

\bigskip
A distinctive feature of string unification is the possibility of
unification by a non-simply-laced group.  This occurs most naturally in
four dimensional type~II string models where the gauge symmetry is realized
by Kac-Moody algebras at different levels.  We investigate the running
coupling constants and the one-loop thresholds for such general models.  As
a specific case, we examine a $\rm SU(3)\times U(1)\times U(1)$ model and
find that the threshold corrections lead to a small $6\%$ increase in the
unification scale.

\Date{May 1992}

\lref\cm{J. A. Casas and C. Mu\~noz, Phys. Lett. {\bf B271} (1991) 85.}
\lref\ant{I. Antoniadis, K. S. Narain and T. R. Taylor, Phys. Lett. {\bf
B267} (1991) 37.}
\lref\anton{I. Antoniadis, Phys. Lett. {\bf B246} (1990) 377.}
\lref\fmtv{S. Ferrara, N. Magnoli, T. R. Taylor and G. Veneziano, Phys.
Lett. {\bf B245} (1990) 409.}
\lref\filq{A. Font, L. E. Ib\'a\~nez, D. L\"ust and F. Quevedo, Phys. Lett.
{\bf B245} (1990) 401.}
\lref\tv{T. R. Taylor and G. Veneziano, Phys. Lett. {\bf B212} (1988) 147.}
\lref\min{J. A. Minahan, Nucl. Phys. {\bf B298} (1988) 36.}
\lref\ekn{J. Ellis, S. Kelley and D. V. Nanopoulos, Phys. Lett. {\bf B260}
(1991) 131.}
\lref\adbf{U. Amaldi, W. de Boer and H. F\"urstenau, Phys. Lett. {\bf B260}
(1991) 447.}
\lref\ll{P. Langacker and M. Luo, Phys. Rev. {\bf D44} (1991) 817.}
\lref\dkl{L. J. Dixon, V. S. Kaplunovsky and J. Louis, Nucl. Phys. {\bf B355}
(1991) 649.}
\lref\ilr{L. E. Ib\'a\~nez, D. L\"ust and G. G. Ross, Phys. Lett. {\bf B272}
(1991) 251.}
\lref\kawai{H. Kawai, D. Lewellen, J. A. Schwartz and H. Tye, Nucl. Phys.
{\bf B299} (1988) 431.}
\lref\abk{I. Antoniadis, C. Bachas and C. Kounnas, Nucl. Phys. {\bf B289}
(1987) 87.}
\lref\kap{V. S. Kaplunovsky, Nucl. Phys. {\bf B307} (1988) 145\semi
Errata, preprint hep-th@xxx/9205068 (May 1992).}
\lref\abbott{L. F. Abbott, Nucl. Phys. {\bf B185} (1981) 189.}
\lref\callan{C. G. Callan, D. Friedan, E. J. Martinec and M. J. Perry, Nucl.
Phys. {\bf B262} (1985) 593.}
\lref\bdg{R. Bluhm, L. Dolan and P. Goddard, Nucl. Phys. {\bf B289} (1987)
364.}
\lref\bdgII{R. Bluhm, L. Dolan and P. Goddard, Nucl. Phys. {\bf B309} (1988)
330.}
\lref\weinberg{S. Weinberg, Phys. Lett. {\bf 91B} (1980) 51.}
\lref\hsw{P. S. Howe, K. S. Stelle and P. C. West, Phys. Lett. {\bf 124B}
(1983) 55.}
\lref\aeln{I. Antoniadis, J. Ellis, R. Lacaze and D. V. Nanopoulos, Phys.
Lett. {\bf B268} (1991) 188.}
\lref\dkv{L. J. Dixon, V. S. Kaplunovsky and C. Vafa, Nucl. Phys. {\bf B294}
(1987) 43.}
\lref\ginsparg{P. Ginsparg, Phys. Lett. {\bf B197} (1987) 139.}

\def\apm{{\alpha'}}
\def\Tr{{\rm Tr}}
\def\zbar{{\bar z}}
\def\wbar{{\bar w}}
\def\qbar{{\bar q}}
\def\Im{{\rm Im}}

\newsec{Introduction}
There has been much excitement over the recent precision electroweak
measurements at LEP and their implication for grand unification.  In
particular, this data shows that with a given normalization, corresponding
to grand unification of the standard model via a simply laced group, the
running coupling constants become equal at a GUT scale of the order
$10^{16}$~GeV\refs{\ekn\adbf{--}\ll}.

In this paper, we are interested in the alternative possibility that the
low-energy physics is unified by a string theory.  At first glance, string
theory, which is perhaps the only known consistent theory of gravity, does
not predict any running of the couplings since it is a finite theory.
However, when we only look at the massless string modes, then they can be
described by an effective field theory.  It is in this context that we talk
about running coupling constants in string theory\refs{\tv,\min}.

Since the string scale is set by the Planck mass, it appears that string
unification would occur at the Planck scale.  For the heterotic string, this
seems to contradict the possible evidence of
unification at $10^{16}$~GeV.  However, explicit calculations of threshold
effects at the string scale show that these contributions may be large
enough to lower the unification scale to a more phenomenologically acceptable
regime.  This is understood because, while a single massive state alone
may not give rise to a large threshold contribution, a string theory has
an infinite tower of massive levels, all of which will contribute.

Although previous investigations of string unification have focused on
heterotic models\refs{\kap\dkl\ilr\aeln\filq\fmtv\ant{--}\cm},
we wish to carry out the analysis in the context of
four dimensional type~II superstring models.  Our reason for doing so is
that one of the goals of string theory is to find a theory with no or few
parameters.  While the heterotic string has a very rich structure, it has
perhaps too much freedom and gives rise to a plethora of possible vacuum
states.  The type~II models are a lot more economical (perhaps too much
so\dkv) and thus furnish a simpler and more constrained ``testing ground''
for string theory.  Even if type~II models turn out to be unrealistic, we
feel it is an ideal testbed for working out techniques of low-energy string
phenomenology because of its relative simplicity.

In order to talk about running coupling constants, we describe the massless
string states by an effective field theory where the massive states have
been integrated out.  We are then concerned with matching this low-energy
theory with the fundamental string theory.  The tree-level relation
between the low-energy coupling constants\ginsparg\ is
$g_i^2=2g_{\rm str}^2/x_i$ where
the factor of 2 arises because we choose a field theory
normalization for the length squared of the longest roots equal to 1
and where $x_i$ is the level of the Kac-Moody algebra (KMA).  Using
$\alpha_i=g_i^2/4\pi$, the
one-loop formula for the running couplings is
\eqn\runc{{1\over\alpha_i(\mu)}={x_i/2\over\alpha_{\rm GUT}}-{b_i\over2\pi}
\log\mu/M_{\rm str}+{\Delta_i\over4\pi}}
where $b_i$ are the $\beta$-function coefficients calculated in the effective
field theory and $\Delta_i$ are threshold corrections arising from the
matching conditions.  $M_{\rm str}$ is an effective string unification
scale and $\alpha_{\rm GUT}$ is an effective coupling
which will be related to the string coupling constant below.

In the following, we calculate the $\Delta_i$ for
general type~II models which naturally include KMA representations with
$x_i>1$.
We use the background field method\abbott\ to calculate the
one-loop effects, and our calculation parallels that of Kaplunovsky\kap.
Finally, we look at a specific example and calculate the thresholds
numerically for an $N=2$ $\rm SU(3)\times U(1)\times U(1)$ model.  We find
that the correction only amounts to a small shift of the unification mass
scale and conclude that thresholds may have a smaller effect in type~II
theories than in heterotic models.

\newsec{Background Field Calculation}
We study the one-loop renormalization of the gauge coupling constants
through the background field method.  This
allows us to quantize the gauge field by treating it as a classical
background in which the strings propagate.  We note that this approach is
different from that of the conventional ``strings in background fields''
work\callan.  The standard work focuses on studying the consistent
propagation of strings in curved space and background gauge fields.  This
is achieved when the string sigma-model $\beta$-functions vanish and the
model is conformally invariant.  For that work, it is sufficient to look at
the string tree level.  On the other hand, we are concerned with the
space-time gauge coupling constant $\beta$-function which we calculate at
the string one-loop level.

Our starting point is the string sigma-model describing the propagation of
strings in a background gauge field.  The string effective action in the
presence of an $A_\mu$ background can be represented as
$\Gamma_{\sigma-{\rm model}}[X^\mu,\psi^\mu;A_\mu]$ where $X^\mu$ and
$\psi^\mu$ are the
string fields.  The background field method\abbott\ tells us that the
effective action for the gauge field is given by $\Gamma[A_\mu]=
\Gamma_{\sigma-{\rm model}}[X^\mu=0,\psi^\mu=0;A_\mu]$ where $\Gamma[A_\mu]
=\int d^4x{\cal L}(A_\mu)$.  In other words, the gauge field effective action
is given by the partition function of the string in the presence of
that gauge field.

At tree level, the effective lagrangian ${\cal L}(A_\mu)$ is simply the
classical Lagrangian
\eqn\treeL{{\cal L}_{\rm tree}(A_\mu)
=\sum_i -{1\over4g_i^2}F_{\mu\nu}^aF^{\mu\nu a}}
where $i$ labels the subgroup.
We are interested in the one-string-loop correction to this tree level
result.  For simplicity, we examine each subgroup independently.  For
subgroup $i$, we turn on a constant background field-strength,
$A_\mu^i(X)=-{1\over2}F_{\mu\nu}X^\nu$, where $F_{\mu\nu}$ is a constant
($A_\mu^j(X)=0$ for $j\ne i$).  If the subgroup is nonabelian, it is
sufficient to shift only one of the gauge fields in the subgroup because of
gauge invariance.  We set all other components in that subgroup to zero so
$F_{\mu\nu}$ above carries no internal gauge index and is abelian.
Although this
background gauge field will induce a background gravitational field via
Einstein's equations, since we are only interested in
the gauge group dependent thresholds, it is sufficient to calculate the
two-point background gauge field amplitude\kap.  The coefficient of $-{1\over4}
F_{\mu\nu}^2$ in this amplitude is then the one-loop correction to the
tree-level gauge coupling $1\over g_i^2$.

The four dimensional type~II string models\refs{\bdg\bdgII\abk{--}\kawai}\ are
constructed out of 20
left-moving (2 space-time and 18 internal) and 20 right-moving light-cone
gauge free fermions.  A particular model is described by a
set of boundary conditions, $\Omega$, for the 40 fermions.  The partition
function is then given as a sum over the sectors (labeled by $\alpha$
and $\beta$) generated by the
set of $\rho_\alpha\equiv(\rho_\alpha;\bar \rho_\alpha) \in\Omega$
\eqn\part{\eqalign{Z&={1\over2^{K+1}}\sum_{\alpha,\beta}c(\alpha,\beta)
\Tr_\alpha[q^{L_0'-\alpha_0}\bar q^{\bar L_0'-\bar\alpha_0}(-1)^{N_\beta}]\cr
&={1\over2^{K+1}}
\sum_{\alpha,\beta}c(\alpha,\beta)
|\eta(q)|^{-24}
\prod_{i=1}^{20}
\left(\vartheta\left[{\rho_\alpha^i\atop\rho_\beta^i}\right](q)\right)^{1/2}
\prod_{i=1}^{20}\left(\vartheta\left[{\bar\rho_\alpha^i\atop\bar\rho_\beta^i}
\right](\bar q)\right)^{1/2}\cr}}
where $\Omega$ is generated by $K+1$ independent vectors.  The
$c(\alpha,\beta)=\delta_\alpha\epsilon(\alpha,\beta)$ are phases for the
$(-1)^{N_\beta}$projections as described in \bdgII.  In the above, the
prime denotes the omission of the bosonic zero mode, $p$.

We work in the operator formalism in order to fix the various normalizations.
The one-loop two-point amplitude contribution to the effective lagrangian
for the $A_\mu^i$
background can be written in this operator language as
\eqn\onel{{\cal L}'(A_\mu^i)={1\over2^{K+1}}\sum_{\alpha,\beta}c(\alpha,\beta)
\int{d^4p\over
(2\pi)^4}\Tr_\alpha[\Delta V^i(1,1)\Delta V^i(1,1)(-1)^{N_\beta}]}
where the sum over sectors corresponds to a generalized GSO
projection\bdgII.  The closed string propagator is
\eqn\prop{\Delta={\apm\over2\pi}\int_{|z|\le1}{dz d\bar z\over|z|^2}
z^{L_0-\alpha_0}\bar z^{\bar L_0-\bar\alpha_0}.}
Although this is written in the (NS,NS) sector where
$\alpha_0=\bar\alpha_0=1/2$, other spin structures are
treated similarly.  In general, $\alpha_0=-1/12 -b/48+d/24$ where $(b,d)$ is
the total number of NS and R left moving fermions respectively in a given
sector, and $\bar\alpha_0$ is similarly defined for right moving fermions.

The $V^i(1,1)$ are conformal weight (1,1) background field vertices.  They are
created from the ordinary type~II vertex (in covariant gauge) for emission
of a gauge boson with state $\tilde b_{-1/2}^a\epsilon\cdot b_{-1/2}|k\rangle$
\eqn\vert{\eqalign{V^a(k,\epsilon,z,\zbar)={}&[\half\sqrt{2\apm}k\cdot
\tilde\psi(z)\tilde \psi^a(z)-{\textstyle{i\over2}}f_{abc}\tilde \psi^b(z)
\tilde \psi^c(z)]\cr
&\epsilon\cdot[i\zbar\bar \partial X_{\rm R}(\zbar)-\half\sqrt{2\apm}
\psi(\zbar) k\cdot\psi(\zbar)]e^{i\sqrt{2\apm}k\cdot X(z,\zbar)}\cr}}
by the substitution $\epsilon_\mu e^{i\sqrt{2\apm}k\cdot X} \to A_\mu(X)$
provided $A_\mu$ solves the equations of motion, $\partial_\mu
F^{\mu\nu}=0$.  For a constant $F_{\mu\nu}$ corresponding to a given
component of subgroup $i$, the resulting background field vertex is
\eqn\bvert{\eqalign{V^i[F_{\mu\nu}](z,\zbar)={\textstyle{i\over4}}F_{\mu\nu}&
\bigl\{
J^a(z)[2X^\mu(z,\zbar)\zbar\bar\partial X_{\rm R}^\nu(\zbar)-\psi^\mu(\zbar)
\psi^\nu(\zbar)]\cr
&-i[\tilde\psi^\mu(z)\tilde \psi^a(z)][\zbar
\bar\partial X_{\rm R}^\nu(\zbar)]\bigr\}\delta^{ai}.\cr}}
Here,
\eqn\Xosc{\eqalign{X^\mu(z,\zbar)&={x^\mu\over\sqrt{2\apm}}
+{\sqrt{2\apm}p^\mu\over4i}(\ln z+\ln\zbar)
+{i\over2}\sum_{n\ne0} {1\over n}\tilde\alpha_n^\mu z^{-n}
+{i\over2}\sum_{n\ne0} {1\over n}\alpha_n^\mu \zbar^{-n}\cr
&={1\over2}\left(X_{\rm L}^\mu(z)+X_{\rm R}^\mu(\zbar)\right)\cr}}
and
\eqn\Jqm{J^a(z)=-{i\over2}f_{abc}\tilde \psi^b(z)\tilde \psi^c(z)}
is the quark model current satisfying a KMA
\eqn\KM{J^a(z)J^b(w)={k_i\delta^{ab}\over(z-w)^2}+{if_{abc}\over z-w}J^c(w)
+\hbox{regular terms}}
with $k_i=C_\psi/2$.  Since the $f_{abc}$ are normalized by
$f_{abc}f_{abe}=C_\psi\delta_{ce}=2\delta_{ce}$, all $k_i=1$.  This value of
$k_i$ is dependent on the root normalization whereas the level of the KMA,
\vadjust{\nobreak}
given by $x_i=2k_i\tilde h_i/C_\psi$ for $\tilde h_i$ the dual Coxeter
number of a given subgroup defined by $\tilde h=C_\psi/\psi^2$ with $\psi^2$
being the length squared of the longest root in the subgroup ($\tilde h=N$
for SU($N$)), is independent of any normalization.
This point is
important when we wish to compare the string theory results where $C_\psi=2$,
corresponding to $\psi_i^2=2/\tilde h_i$,
with the conventional field theory results which are given in a normalization
with all roots $\psi^2_i=1$.
Written in this manner, the first line in \bvert\ is identical to the
heterotic background vertex\kap\ except that $J^a$, given by \Jqm, is no
longer a level 1 representation.  The additional term in \bvert\ is due to
the gauge group being generated by a super KMA in the present
case of the type~II string.
The $\delta^{ai}$ in \bvert\ denotes that we only turn on a single component
of the background gauge field in each subgroup.  Because of gauge
invariance, it is unimportant which specific component we choose.

Using standard operator methods, we rewrite \onel\ as
\eqn\leff{\eqalign{{\cal L}'(A_\mu^i)={1\over2^{K+1}}\sum_{\alpha,\beta}
c(\alpha,\beta)
(2\apm\pi)^2&\int{d^4p\over(2\pi)^4}\int_\Gamma d^2\tau
\int\limits_{0\le\Im\nu\le\Im\tau}\!\!\!\!\!\!\!\!\!d^2\nu\cr
&\quad\quad\Tr_\alpha[V^i(z,\zbar)V^i(1,1)
q^{L_0-\alpha_0}\qbar^{\bar L_0-\bar\alpha_0}(-1)^{N_\beta}]\cr}}
where $z=e^{2\pi i\nu}$ and $q=e^{2\pi i\tau}$ and we have restricted the
modular integral over the fundamental region $\Gamma$.  The operator trace
gives the two-point correlator $\langle VV\rangle$ on the torus
\eqn\brkt{\eqalign{
\langle V^i(z,\zbar)V^i(w,\wbar)\rangle_{\alpha,\beta}&\equiv
{
\Tr_\alpha[V^i(z,\zbar)V^i(w,\wbar)
q^{L_0-\alpha_0}\qbar^{\bar L_0-\bar\alpha_0}(-1)^{N_\beta}]
\over
\Tr_\alpha[ q^{L_0-\alpha_0}\qbar^{\bar L_0-\bar\alpha_0}(-1)^{N_\beta}]
}\cr
&=-{1\over16}F_{\mu\nu}F_{\rho\sigma}\biggl\{
\langle J^i(z)J^i(w)\rangle\cr
&\qquad\qquad\qquad\quad\times
\biggl[\langle2X^\mu(z,\zbar)\zbar\bar\partial X_{\rm R}^\nu(\zbar)
2X^\rho(w,\wbar)\wbar\bar\partial X_{\rm R}^\sigma(\wbar)\rangle\cr
&\qquad\qquad\qquad\qquad\quad
+\langle\psi^\mu(\zbar)\psi^\nu(\zbar)\psi^\rho(\wbar)\psi^\sigma(\wbar)
\rangle\biggr]\cr
&\qquad\qquad\qquad
-\langle\tilde\psi^\mu(z)\tilde\psi^i(z)\tilde\psi^\rho(w)\tilde\psi^i(w)
\rangle\langle\zbar\bar\partial X_{\rm R}^\nu(\zbar)\wbar
\bar\partial X_{\rm R}^\sigma(\wbar)\rangle\biggr\}.\cr}}
After
dropping total derivatives, we find that the last term above, which comes
from the additional term in \bvert, does
not contribute to ${\cal L}'(A_\mu^i)$.  Performing the $p$ integration then
gives
\eqn\vvcor{\eqalign{{\cal L}'(A_\mu^i)={1\over4}F_{\mu\nu}^2&{1\over16\pi^2}
{1\over2^{K+1}}\sum_{\alpha,\beta}c(\alpha,\beta)\int_\Gamma
{d^2\tau\over\tau_2}\int{d^2\nu\over\tau_2}\langle J^i(z)J^i(1)
\rangle_{\alpha,\beta}\cr
&\times2\left[\langle\psi(\zbar)\psi(1)\rangle^2_{\alpha,\beta}
-\langle X_{\rm R}(\zbar)\zbar\bar\partial X_{\rm R}(1)\rangle^2\right]
\Tr_\alpha[q^{L_0'-\alpha_0}\bar q^{\bar L_0'-\bar\alpha_0}(-1)^{N_\beta}]\cr}}
which is similar to the expression for the heterotic case where we have
corrected several numerical factors in \kap.

Using \KM, we find that
\eqn\JJcor{\langle J^i(z)J^i(1)\rangle=-k_i\left(z{\partial\over\partial z}
\right)^2\log\vartheta_1(z,q)+\langle J^i_0 J^i_0\rangle.}
As pointed out in \kap, the first term yields a gauge group independent
(apart from the $k_i$) term which can be combined with the induced
gravitational background term.  We thus drop this piece and concentrate on
the group dependent charges $\langle J^i_0 J^i_0\rangle$.  For the
space-time correlators, both $\langle\psi\psi\rangle^2$ and $\langle X_{\rm
R} \zbar\bar\partial X_{\rm R}\rangle^2$ have double poles in the $\nu$
plane and are
related to the Weierstrass $\cal P$ function.  As a result, the difference is
finite, so the $\nu$ integral can be performed to get
\eqn\stcor{\eqalign{{\cal L}'(A_\mu^i)={1\over4}F_{\mu\nu}^2 {1\over16
\pi^2}&{1\over2^{K+1}}
\sum_{\alpha,\beta}c(\alpha,\beta)\int_\Gamma{d^2\tau\over\tau_2}
2\langle J^i_0J^i_0\rangle_{\alpha,\beta}\cr
&\times2\qbar{d\over d\qbar}\log
\left(
{\vartheta\left[\bar\rho^1_\alpha\atop\bar\rho^1_\beta\right](\qbar)
\bigg/ 
\eta(\qbar)}
\right)
\Tr_\alpha[q^{L_0'-\alpha_0}\qbar^{\bar L_0'-\bar\alpha_0}
(-1)^{N_\beta}].\cr}}
Since all space-time fermions are moded identically, we have arbitrarily
picked the boundary condition of the first fermion.

At this stage, the expression is equivalent to that for the heterotic
string.  However, for the type~II string, we can now use the explicit
realization \Jqm\ of the $J^a(z)$ in terms of free fermions in order to
evaluate
$\langle J^i_0J^i_0\rangle$ explicitly.  For a symmetric subgroup, both
fermions generating $J_0^a$ are moded identically, and the result is
\eqn\JJtwo{\langle J^i_0J^i_0\rangle=\half f^i{}_{cd}f^i{}_{cd}\,2q{d\over dq}
\log\vartheta\left[{\rho^c_\alpha\atop\rho^c_\beta}\right](q)}
(no sum on $i$).  The spin structure is that of fermion $c$ (or equivalently
$d$).  Using the form of the partition function, \part, we finally arrive at
the expression
\eqn\lfin{{\cal L}'(A_\mu^i)=-{1\over4}F_{\mu\nu}^2{1\over16\pi^2}\int_\Gamma
{d^2\tau\over\tau_2}2B_i(q,\qbar)}
where
\eqn\bqq{\eqalign{B_i(q,\qbar)=-&{1\over2^{K+1}}\sum_{\alpha,\beta}
c(\alpha,\beta)
|\eta(q)|^{-23}\vartheta\left[{\rho^1_\alpha\atop\rho^1_\beta}\right](q)
\;2\qbar{d\over
d\qbar}\left(\vartheta\left[{\bar\rho^1_\alpha\atop\bar\rho^1_\beta}
\right](\qbar)\bigg/\eta(\qbar)\right)\cr
&\times\prod_{j=3}^{20}\left(\vartheta\left[{\rho^j_\alpha\atop\rho^j_\beta}
\right](q)\right)^{1/2}\prod_{j=3}^{20}\left(\vartheta
\left[{\bar\rho^j_\alpha\atop
\bar\rho^j_\beta}\right](\qbar)\right)^{1/2}\half
f^i{}_{cd}f^i{}_{cd}\,2q{d\over dq}
\log\vartheta\left[{\rho^c_\alpha\atop\rho^c_\beta}\right](q).\cr}}

Recall that the one-loop renormalization to $1\over g_i^2$ is the coefficient
of $-{1\over4}F_{\mu\nu}^2$.  We now want to match this string theory result
to that of the low-energy effective field theory of the massless particles
in four dimensions\refs{\kap,\weinberg}.  In order to match the
normalizations for nonabelian groups,
we have to convert from the type~II string normalization of $C_\psi=2$ to the
field theory convention of $\psi^2_{\rm FT}=1$.  This gives an overall factor
of
$\psi^2_{\rm FT}/\psi^2_{\rm str}=\tilde h_i/2=x_i/2$ for subgroup $i$.
For abelian groups, the level is undefined.  In this case, we can choose
$x_{\rm U(1)}=2$ which corresponds to using the same normalization for the
U(1) charges in field theory and string theory.
Equating the bare couplings then gives
\eqn\bare{{16\pi^2\over g_i^2}+b_i\int_0^\infty{dt\over t}C_\Lambda(t)
={x_i\over2}\left[{16\pi^2\over g_{\rm str}^2}+\int_\Gamma{d^2\tau\over\tau_2}2
B_i(q,\qbar)+Y\right]}
where $C_\Lambda(t)$ is a field theory ultraviolet cutoff.  $Y$ is the
gauge group independent terms that we have not evaluated.
The effect of $Y$ is to shift the string coupling constant and can be
absorbed by a redefinition of the unified coupling
constant\foot{Strictly speaking, modular anomalies arise in the above
expressions because the first term of \JJcor\ has double poles on the
world-sheet and must be regulated\dkl.  However, the differences
between the gauge couplings, which are our main interest, are indeed
modular invariant.  As a result, $\alpha_{\rm GUT}$ itself should only
be viewed as a formal expression.}
\eqn\agut{{1\over\alpha_{\rm GUT}}={4\pi\over g_{\rm str}^2}+{Y\over4\pi}.}
The $b_i$ are the
field theory $\beta$-functions and agree with the string calculation
\eqn\bfun{b_i=\lim_{q\to0}x_iB_i(q,\qbar)=-{11\over3}\Tr_{\rm V}(Q_i^2)
+{2\over3}\Tr_{\rm F}(Q_i^2)+{1\over6}\Tr_{\rm S}(Q_i^2)}
(in field theory normalization).  Here, the traces are over two-component
fermions and real scalers.

Using $\alpha_i=g_i^2/4\pi$ and following \kap\ in converting \bare\ into an
expression for the $\overline{\rm DR}$ couplings, we finally arrive
at the matching formula, \runc, for the running couplings
\eqn\run{{1\over\alpha_i(\mu)}={x_i/2\over\alpha_{\rm GUT}}-{b_i\over2\pi}
\log\mu/M_{\rm str}+{\Delta_i\over4\pi}.}
The thresholds are
\eqn\thr{\Delta_i=\int_\Gamma{d^2\tau\over\tau_2}[x_iB_i(q,\qbar)-b_i]}
and the string unification scale is
\eqn\mstr{M_{\rm str}^2={2e^{(1-\gamma)}\over\sqrt{27}\pi\apm}}
where $\gamma$ is the Euler constant.  Because of the string relation,
$\kappa=\half g_{\rm st}\sqrt{2\apm}$, the scale $M_{\rm str}$ can be
related in the $\overline{\rm DR}$ scheme to the
string coupling constant with the result\refs{\kap,\ilr}
\eqn\msmpl{M_{\rm str}=0.7\times g_{\rm str}\times 10^{18}\rm\ GeV.}
Such a constraint on the unification scale is an additional feature of
string unification that is not found in conventional grand unification.

\newsec{An $N=2$ model}
For an explicit example of how this formalism works, we now turn to specific
type~II models.  For the original $N=4$ models with dimension 18 gauge
groups $\rm SU(2)^6$, $\rm SU(4)\times SU(2)$ or $\rm SU(3)\times SO(5)$, eight
right-moving fermions are moded identically, and the $\beta$-functions and
thresholds both vanish by $N=4$ supersymmetry as in \kap.  This result is
easily interpreted by realizing that $N=4$ gives a finite supersymmetric gauge
field theory.

While as of yet there are no realistic type~II string models, the $N=2$
$\rm SU(3)\times U(1)\times U(1)$ model of \bdgII\ has some interesting
features.  This model is given by $K=2$ and is generated by the basis\bdgII
\eqn\bas{\eqalign{
\rho_{b_0}&=\left((1)^{12},(1)^8;(1)^4,(1)^8,(1)^4,(1)^4\right)\cr
\rho_{b_1}&=\left((1)^{12},(0)^8;(1)^4,(1)^8,(0)^4,(0)^4\right)\cr
\rho_{b_2}&=\left((1)^{12},(1)^8;(0)^4,(1)^8,(1)^4,(0)^4\right).\cr}}

The massless states come from the sectors with four or less Ramond fermions on
each side.  These are given by the boundary conditions $\emptyset$, $b_0b_1$,
$b_0b_2$ and $b_1b_2$.  The
states from the bosonic and the fermionic sectors combine to form an $N=2$
supergravity multiplet with helicities $(\pm2,2(\pm3/2),\pm1)$, an $N=2$
super-Yang-Mills multiplet with helicities $(\pm1,2(\pm1/2),2(0))$ in a
singlet and adjoint representation representation of the gauge group, and
$N=2$ matter with helicities $(2(\pm1/2),4(0))$ in the $\rm SU(3)\times U(1)
\times U(1)$ representations
\eqn\reps{\eqalign{[{\bf 1},0,&\pm1]\oplus[{\bf 3},-1/\sqrt{3},0]\oplus
[\bar{\bf 3},1/\sqrt{3},0]\cr
&\oplus[{\bf 1},\sqrt{3}/2,\pm\half]\oplus[{\bf 1},-\sqrt{3}/2,\pm\half]\oplus
[{\bf 3},1/(2\sqrt{3}),\pm\half]\oplus[\bar{\bf 3},-1/(2\sqrt{3}),\pm\half]
.\cr}}

This model arises from the $N=4$ $\rm SU(4)\times SU(2)$ model by
considering the symmetric subgroups $\rm SU(3)\times U(1)$ and U(1) of SU(4)
and SU(2) respectively.  As a result, the SU(3) is in a level 4
representation of the KMA so $x_i=(4,2,2)$ where
we have used the natural normalization that arises from the string
theory for the U(1) factors.
The field theory $\beta$-functions, \bfun, are easily calculated
from the massless spectrum.  The result is $b_i=(0,12,12)$.  Since the SU(3)
$\beta$-function vanishes at one-loop, it actually vanishes at all orders
because this is a sufficient condition for finiteness of the $N=2$
super-Yang-Mills coupled to $N=2$ matter theory\hsw.

Using the explicit form of the $\rm SU(4)\times SU(2)$ structure functions,
we see that the fermions generating the gauge group are either both moded
like the first twelve or the last eight left-moving world-sheet fermions.  As
a result,
we find that $B_{\rm SU(3)}(q,\qbar)=3/4B_A(q,\qbar)+1/4B_\alpha(q,\qbar)$
and $B_{\rm U(1)}(q,\qbar)=B_\alpha(q,\qbar)$ where $A$ refers to the first,
and $\alpha$ refers to the last set of boundary conditions.  Both U(1)
$\beta$-functions and thresholds are identical at the string level.

In order to calculate $B_A(q,\qbar)$ and $B_\alpha(q,\qbar)$, we expand the
$\vartheta$-functions in \bqq\ in terms of $q$ and perform a numerical
integration over the fundamental region.  Since $\tau_2\ge\sqrt{3}/2$ in the
fundamental region,
$|q|\le e^{-\sqrt{3}\pi}\approx 0.0043$, so the series expansions converge
rapidly.  While only sectors with massless particles
contribute to the $\beta$-functions, all sectors contribute to the massive
string thresholds.  We numerically evaluate each sector independently taking
care to combine differences of $\vartheta$-functions analytically to
minimize numerical errors in the limit $\tau_2\to\infty$.  The actual
double integration is performed by open Romberg integrations in $\tau_1$ and
$\tau_2$.  Because $B_i(-\tau_1,\tau_2)=B_i(\tau_1,\tau_2)^*$ where $*$
denotes the complex conjugate, it is sufficient to integrate twice the real
part of $B_i$ over half of the fundamental region.  The thresholds are, of
course, real.  The result of
this numerical integration is the thresholds $\Delta_i=(-2.35,0.144,0.144)$.

Since there are only two independent running couplings, we define the
effective unification scale, $M_{\rm U}$, to be where the string theory
normalized couplings meet, i.e.~$1/\alpha_i(M_{\rm U})=(x_i/2)
/\alpha_{\rm GUT}$.
The difference, $\Delta_{\rm SU(3)}-\Delta_{\rm U(1)}=-1.32$ (in string
theory normalization), then corresponds to an increase of the effective
unification scale over $M_{\rm str}$ of
\eqn\Mincr{M_{\rm U}/M_{\rm str}=\exp\left({1\over2}{\Delta_{\rm SU(3)}
-\Delta_{\rm U(1)}\over b_{\rm SU(3)}-b_{\rm U(1)}}\right)=
\exp\left({1.32\over24}\right)=1.06\;.}
This $6\%$ increase is rather small compared to estimates made for
some heterotic models\aeln.  It is difficult to interpret this
result physically since this is not a particularly realistic model.
However, this leads us to believe that thresholds may play a smaller role in
type~II string unification than it does in the heterotic case.

\newsec{Conclusion}
In general orbifold models of the heterotic string, any $N=4$ and $N=2$
orbifold sectors will lead to compactification moduli.  In these
models, the thresholds, $\Delta_i$, can be expressed roughly in terms of
the moduli from the $N=2$ sectors as\refs{\dkl,\filq,\fmtv}
\eqn\modi{\Delta_i(T,\bar T)\approx-b_i\log[(T+\bar T)|\eta(T)|^4]}
where the complex modulus, $T$, is composed of the radius and an internal
axion field of the 2-torus fixed in an $N=2$ sector.  This expression is
pretty much fixed by the requirement of target space modular invariance.  By
absorbing this threshold into a redefinition of $M_{\rm U}$, we can
interpret \modi\ as indicating that the effective coupling constants, $g_i$,
only start running below a radius dependent scale,
$M_{\rm U}$\refs{\tv,\anton}.

On the other hand, for these type~II  models, all moduli are fixed to be at
the point where the gauge symmetry is enlarged.  As a result, there are no
free parameters in the thresholds that can be adjusted to give large
corrections.  This feature of type~II
models would seem to indicate that the effective unification scale cannot
easily be lowered much below \msmpl.  However, this should not be viewed as
a drawback since the couplings are no longer required to meet at a point at
the unification scale.

We have examined the running of the effective coupling constants and
one-loop threshold effects in four-dimensional type~II string theory as a
possible first step towards relating strings to low-energy phenomenology.
Our interest in studying the type~II models is that it provides a simpler
framework than the heterotic string where the fundamental ideas of
unification can be understood.  Work is in progress to develop more realistic
type~II models.  When this is completed, it would be a simple matter to
compute the gauge $\beta$-functions, $b_i$, from the massless particle
spectrum and the thresholds, $\Delta_i$, from \thr.  This would then be
sufficient for predicting the low-energy parameters $\sin^2\theta_W$ and
$\alpha_s$ at the $Z^0$ mass which may eventually lead to some testable
results of string theory.

\bigskip
This work is supported in part by the U.S.~Department of Energy under Grant
No.~DE-FG05-85ER-40219.

\listrefs
\bye